\documentclass[pre,onecolumn]{revtex4}
\usepackage{amssymb}
\usepackage{amsmath}
\usepackage{graphicx}
\begin{document}
\title{Nonequilibrium Gaussian Distribution for Coupled Linear Langevin Equations}
\author{Hidetsugu Sakaguchi}
\address{Interdisciplinary Graduate School of Engineering Sciences, \\
Kyushu University, Kasuga, Fukuoka 816-8580, Japan}
\begin{abstract}
Explicit forms of nonequilibrium Gaussian distributions and heat flows are obtained for the Fokker-Planck equation corresponding to the coupled linear Langevin equations of two and three variables.  
\end{abstract}
\maketitle
The statistical mechanics can be expressed with the Langevin equation and Fokker-Planck equation~\cite{Kubo}. 
Coupled linear Langevin equations have been studied in equilibrium and nonequilibrium statistical mechanics for interacting particles~\cite{Wang}. Non-reciprocity is a recent topic in several nonequilibrium systems~\cite{Scheibner}. 
General theory of the Langevin equations and Fokker-Planck equations and the relation to the Gaussian distributions are discussed in the book by Gardiner~\cite{Gardiner}. In this short note, we show the explicit form of the Gaussian distribution and heat flows for coupled linear Langevin equations under the two or three heat reservoirs with different temperatures as the simplest example of linear nonequilibrium problems. The model equations are
\begin{eqnarray}
\frac{dx}{dt}&=&-ax+by+\xi_x(t),\nonumber\\
\frac{dy}{dt}&=&cx-dy+\xi_y(t),
\end{eqnarray}
where $\xi_x(t)$ and $\xi_y$ are the Gaussian white noises satisfying $\langle \xi_x(t)\xi_x(t^{\prime})\rangle=2T_x\delta(t-t^{\prime})$ and $\langle \xi_y(t)\xi_y(t^{\prime})=2T_y\delta(t-t^{\prime})\rangle$. The thermal equilibrium is not assumed. The corresponding Fokker Planck equation is 
\begin{equation}
\frac{\partial P}{\partial t}=\frac{\partial}{\partial x}\{(ax-by)P\}+\frac{\partial}{\partial y}\{(-cx+dy)P\}+T_x\frac{\partial^2P}{\partial x^2}+T_y\frac{\partial^2P}{\partial y^2}. \label{fp}
\end{equation} 
Since the two variables $x$ and $y$ are generated from the linear combination of Gaussian noises, the stationary probability distribution is expected to obey the Gaussian distribution: 
\begin{equation}
P=\frac{\pi}{\sqrt{\alpha\gamma-\beta^2/4}}e^{-\alpha x^2-\beta x y-\gamma y^2}. \label{gau}
\end{equation} 
Substitution of Eq.~(\ref{gau}) into Eq.~(\ref{fp}) yields
\begin{eqnarray}
a+d-2\alpha T_x-2\gamma T_y&=&0,\nonumber\\
-2a\alpha+4T_x\alpha^2+c\beta+T_y\beta^2&=&0,\nonumber\\
-a\beta+2b\alpha+4T_x\alpha\beta-d\beta+2c\gamma+4T_y\gamma\beta&=&0,\nonumber\\b\beta+T_x\beta^2-2d\gamma+4T_y\gamma^2&=&0.\nonumber
\end{eqnarray}
The three variables $\alpha$, $\beta$, and $\gamma$ are explicitly expressed as \begin{eqnarray}
\alpha&=&\frac{(a+d)(T_ya^2+T_yad+T_xc^2-T_ybc)}{2T_xT_y(a+d)^2+2(T_xc-T_yb)^2},\nonumber\\
\gamma&=&\frac{(a+d)(T_yb^2-T_xbc+T_xad+T_xd^2)}{2T_xT_y(a+d)^2+2(T_xc-T_yb)^2},\nonumber\\
\beta&=&-\frac{2b\alpha+2c\gamma}{a+d}.
\end{eqnarray}
For the Guassian distribution (\ref{gau}), 
\begin{equation}
\langle x^2\rangle=\frac{2\gamma}{4\alpha\gamma-\beta^2},\;\;\langle y^2\rangle=\frac{2\alpha}{4\alpha\gamma-\beta^2},\;\;\langle xy\rangle=-\frac{\beta}{4\alpha\gamma-\beta^2}.
\end{equation}
The probability flows $J_x$ and $J_y$ are expressed as
\begin{eqnarray}
J_x&=&(ax-by)P+T_x\frac{\partial P}{\partial x},\nonumber\\
J_y&=&(-cx+dy)P+T_y\frac{\partial P}{\partial y}
\end{eqnarray}
If $J_x=J_y=0$, the detailed balance condition is satisfied, and the thermal equilibrium state is realized. The equilibrium distribution is expressed as 
\begin{equation}
P\propto \exp\{(-ax^2+2bxy-(dc/b)y^2)/(2T_x)\}
\end{equation}
where $T_y$ needs to satisfy $T_y=(c/b)T_x$. 
In a system of two variables, there is a temperature $T_y$ where the detailed balance condition is satisfied even if the coupling coefficients are not reciprocal $c\ne b$.  
If $bT_y\ne cT_x$, the nonequilibrium stationary state is realized, and the probability flows exist. 

If $a=0$, $b=1$, $c=-k$, $d=\eta$, and $T_x=0$, Eq.~(1) is 
\[\frac{dx}{dt}=v,\; \frac{dv}{dt}=-kx-\eta y+\xi_y(t),\]
The stationary distribution is $P\propto e^{-(kx^2+y^2)/(2T)}$ where $T=T_y/\eta$. 
If $y$ is interpreted as the velocity, this is the Boltzmann distribution for the total energy, $E=(1/2)kx^2+(1/2)v^2$ for the temperature $T$, mass $m=1$, and the Boltzmann constant $k_{\rm B}=1$. Since $J_x$ and $J_y$ are not zero, the detailed balance condition is not satisfied for the Fokker-Planck equation of $x$ and $y$. 
 
In the reciprocal system $b=c$, the heat flows into the two heat reservoirs are calculated as~\cite{Sekimoto}
\begin{eqnarray}
Q_x&=&\langle (ax-by)J_x\rangle\nonumber\\
&=&\frac{2a^2\gamma+2ab\beta+2b^2\alpha-4T_xa\alpha\gamma+T_xa\beta^2}{4\alpha\gamma-\beta^2},\nonumber\\
Q_y&=&\langle (-cx+dy)J_y\rangle\nonumber\\
&=&\frac{2c^2\gamma+2cd\beta+2d^2\alpha+T_yd\beta^2-4T_yd\alpha\gamma}{4\alpha\gamma-\beta^2}.
\end{eqnarray}
For $a=d=k+l$ and $b=c=l$, Eq.~(1) is $dx/dt=-kx+l(y-x)+\xi_x(t)$ and $dy/dt=-ky+l((x-y)+\xi_y(t)$, and the heat flow $Q=Q_x$ is explicitly expressed as
\begin{equation}
Q=\frac{Q_x-Q_y}{2}=\frac{l^2}{2(k+l)}(T_y-T_x).
\end{equation}

Next, we consider the coupled linear Langevin equations with three variables. The Langevin equations are expressed as
\begin{eqnarray}
\frac{dx}{dt}&=&a_{xx}x+a_{xy}y+a_{az}z+\xi_x(t),\nonumber\\
\frac{dy}{dt}&=&a_{yx}x+a_{yy}y+a_{yz}z+\xi_y(t),\nonumber\\
\frac{dz}{dt}&=&a_{zx}x+a_{zy}y+a_{zz}z+\xi_z(t), \label{lan}
\end{eqnarray}
where $\xi_x(t)$, $\xi_y$, and $\xi_z(y)$ are the Gaussian white noises satisfying $\langle \xi_x(t)\xi_x(t^{\prime})\rangle=2T_x\delta(t-t^{\prime})$, $\langle \xi_y(t)\xi_y(t^{\prime})=2T_y\delta(t-t^{\prime})\rangle$, and $\langle \xi_z(t)\xi_z(t^{\prime})=2T_z\delta(t-t^{\prime})\rangle$
The stationary solution of the corresponding Fokker-Planck equation is expressed as 
\begin{equation}
P\propto \exp(-b_{xx}x^2-b_{yy}y^2-b_{zz}z^2-b_{xy}xy^b_{yz}yz-b_{zx}zx). \label{gau2}
\end{equation}
the coefficients $b_{xx},\cdots, b_{zz}$ satisfy
\begin{eqnarray}
& &-a_{xx}-a_{yy}-a_{zz}-2(T_xb_{xx}+T_yb_{yy}+T_zb_{zz})=0,\nonumber\\
& &2a_{xx}b_{xx}+a_{yx}b_{xy}+a_{zx}b_{zx}+4T_xb_{xx}^2+T_yb_{xy}^2+T_zb_{zx}^2=0,\nonumber\\
& &a_{xy}b_{xy}+2a_{yy}b_{yy}+a_{zy}b_{yz}+T_xb_{xy}^2+4T_yb_{yy}^2+T_zb_{yz}^2=0,\nonumber\\
& &a_{xz}b_{zx}+a_{yz}b_{yz}+2a_{zz}b_{zz}+T_xb_{zx}^2+T_yb_{yz}^2+4T_zb_{zz}^2=0,\nonumber\\
& &2a_{xy}b_{xx}+a_{xx}b_{xy}+2a_{yx}b_{yy}+a_{yy}b_{xy}+a_{zx}b_{yz}+a_{zy}b_{zx}+4T_xb_{xx}b_{xy}+4T_yb_{xy}b_{yy}+2T_zb_{zx}b_{yz}=0,\nonumber\\
& &a_{xy}b_{zx}+a_{xz}b_{xy}+a_{yy}b_{yz}+2a_{yz}b_{yy}+2a_{zy}b_{zz}+a_{zz}b_{yz}+2T_xb_{xy}b_{zx}+4T_yb_{yy}b_{yz}+4T_zb_{yz}b_{zz}=0,\nonumber\\
& &a_{xx}b_{zx}+2a_{xz}b_{xx}+a_{yx}b_{yz}+a_{yz}b_{xy}+2a_{zx}b_{zz}+a_{zz}b_{zx}+4T_xb_{xx}b_{zx}+2T_yb_{xy}b_{yz}+4T_zb_{zx}b_{zz}=0.
\end{eqnarray}
If the coupling constants are reciprocal, i.e., $a_{xy}=a_{yx}$, $a_{yz}=a_{zy}$, and $a_{zx}=a_{xz}$, there is a potential energy, $U=(1/2)a_{xx}x^2+(1/2)a_{yy}y^2+(1/2)a_{zz}z^2+a_{xy}xy+a_{yz}yz+a_{zx}zx$, and the thermal equilibrium distribution is given by $P\propto \exp(-U/T)$ at $T_x=T_y=T_z=T$. If the coupling constants are not reciprocal, the conservativity is not generally satisfied in three-variable systems~\cite{Kolotinskii}. Even in the non-reciprocal system, the stationary Gaussian distribution is obtained.
For $a_{xx}=a_{yy}=a_{zz}=-3$, $a_{xy}=a_{xz}=a_{yz}=a_{yz}=a_{zx}=1$, $a_{zy}=0.2$, $T_x=1$, $T_y=2$, and $T_z=4$, $b_{xx}=1.2780$, $b_{yy}=0.75509$, $b_{zz}=0.4280$, $b_{xy}=-0.8690$, $b_{yz}=-0.1713$, and $b_{zx}=-0.4979$ were numerically obtained for the coupled nonlinear equations (12).  
Figure 1(a) compares the stationary solution of the Fokker Planck equation corresponding to the Langevin equation Eq.~(\ref{lan}) obtained by direct numerical simulation and the Gaussian distribution Eq.~(\ref{gau2}) along the diagonal $x=y=z$. The Gaussian distribution is obtained even in the nonequilibrium stationary state. 

\begin{figure}[h]
\begin{center}
\includegraphics[height=3.5cm]{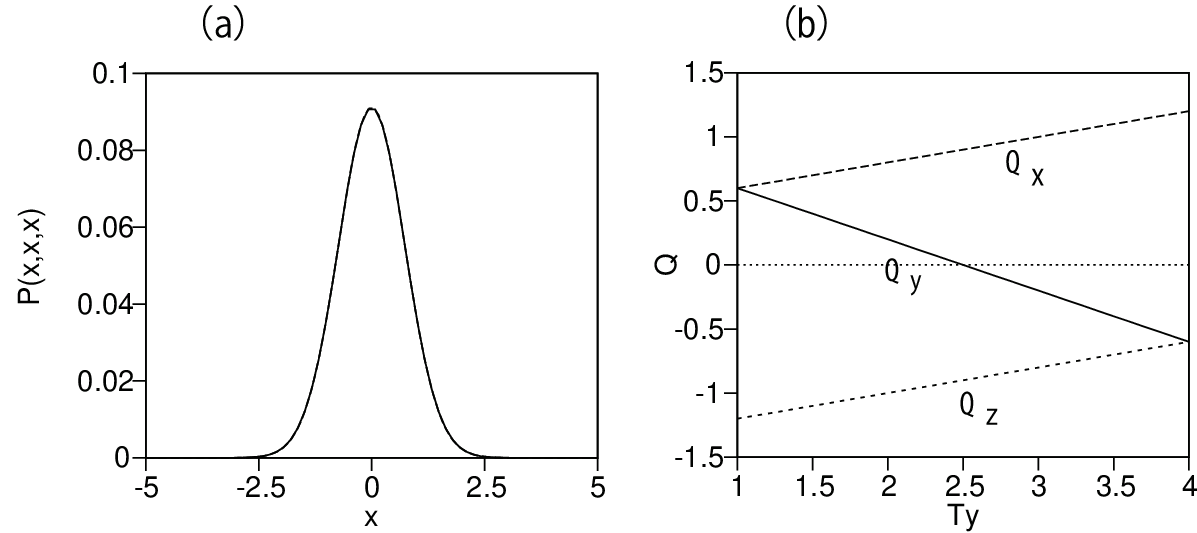}
\end{center}
\caption{(a) Stationary probability distribution (solid line) of the Fokker Planck equation obtained by direct numerical simulation and the Gaussian distribution (dashed line) for $a_{xx}=a_{yy}=a_{zz}=-3$, $a_{xy}=a_{xz}=a_{yz}=a_{yz}=a_{zx}=1$, $a_{zy}=0.2$, $T_x=1$, $T_y=2$, and $T_z=4$. (b) Heat flows $Q_x$, $Q_y$, and $Q_z$ as a function of $T_y$ for $T_x=1$ and $T_z=4$.}
\label{f1}
\end{figure}
In the reciprocal system, the heat flows into the three reservoirs can be calculated from 
\begin{eqnarray}
Q_x&=&\langle(-a_{xx}x-a_{xy}y-a_{xz}z)\frac{\partial J_x}{\partial x}\rangle,\nonumber\\
Q_y&=&\langle(-a_{yx}x-a_{yy}y-a_{yz}z)\frac{\partial J_y}{\partial y}\rangle,\nonumber\\
Q_z&=&\langle(-a_{zx}x-a_{zy}y-a_{zz}z)\frac{\partial J_z}{\partial z}\rangle.
\end{eqnarray}
Here, the averages with respect to the Gaussian distribution can be given by  
\begin{eqnarray}
\langle x^2\rangle&=&\frac{4b_{yy}b_{zz}-b_{yz}^2}{8S},\langle y^2\rangle=\frac{4b_{xx}b_{zz}-b_{zx}^2}{8S},\nonumber\\
\langle z^2\rangle&=&\frac{4b_{xx}b_{yy}-b_{xy}^2}{8S},\langle xy\rangle=\frac{b_{yz}b_{zx}-2b_{zz}b_{xy}}{8S},\nonumber\\
\langle yz\rangle&=&\frac{b_{xy}b_{zx}-2b_{xx}b_{yz}}{8S},\langle zx \rangle=\frac{b_{xy}b_{yz}-2b_{yy}b_{zx}}{8S},\nonumber
\end{eqnarray}
where 
\[S=b_{xx}b_{yy}b_{zz}+(b_{xy}b_{yz}b_{zx}-b_{xx}b_{yz}^2-b_{yy}b_{zx}^2-b_{zz}b_{xy}^2)/4.\]
For example, the heat flows  $Q_x,Q_y$ and $Q_z$ are evaluated as $Q_x=0.8$, $Q_y=0.2$, and $Q_z=-1.0$ for $a_{xx}=a_{yy}=a_{zz}=-3$, $a_{xy}=a_{xz}=a_{yz}=a_{yz}=a_{zx}=a_{zy}=1$, $T_x=1$, $T_y=2$, and $T_z=4$. $Q_x+Q_y+Q_z=0$ is always satisfied.  Figure 1(b) shows $Q_x$, $Q_y$, and $Q_z$ as a function of $T_y$ for $T_x=1$, $T_z=4$,  $a_{xx}=a_{yy}=a_{zz}=-3$, and $a_{xy}=a_{xz}=a_{yz}=a_{yz}=a_{zx}=a_{zy}=1$. $Q_x$, $Q_y$, and $Q_z$ satisfy $Q_x=0.2(T_z-T_x)+0.2(T_y-T_x)$, $Q_y=0.2(T_z-T_y)+0.2(T_x-T_y)$, and $Q_z=0.2(T_x-T_z)+0.2(T_y-T_z)$. The heat flows are proportional to the temperature differences in the coupled linear systems.  

We have shown the Gaussian distributions and heat flows for the coupled linear Langevin equations under the two or three heat reservoirs with different temperatures as one of the simplest linear nonequilibrium problems.  

\end{document}